\documentclass{aastex}
\usepackage{rotating}
\usepackage{atbegshi}
\AtBeginDocument{\AtBeginShipoutNext{\AtBeginShipoutDiscard}}

\begin{document}
\title{Modelling extragalactic extinction through \\ gamma-ray burst afterglows}
\author{Alberto Zonca}
\email{azonca@oa-cagliari.inaf.it}
\affil{INAF\textendash Osservatorio Astronomico di Cagliari, Via della Scienza 5, I-09047 Selargius, Italy} 
\author{Cesare Cecchi-Pestellini}
\email{cecchi-pestellini@astropa.inaf.it}
\affil{INAF\textendash Osservatorio Astronomico di Palermo, P.za Parlamento 1, I-90134 Palermo, Italy} 
\author{Giacomo Mulas, Silvia Casu, Giambattista Aresu}
\affil{INAF\textendash Osservatorio Astronomico di Cagliari, Via della Scienza 5, I-09047 Selargius, Italy \\ \phantom{A}} 
\email{gmulas@oa-cagliari.inaf.it \\ silvia@oa-cagliari.inaf.it \\   garesu@oa-cagliari.inaf.it}

\date{Accepted. Received; in original form \\ \phantom{A}}
\begin{abstract}
We analyze extragalactic extinction profiles derived through gamma-ray burst afterglows, using a dust model specifically constructed on the assumption that dust grains are not immutable but respond time-dependently to the local physics. Such a model includes core-mantle spherical particles of mixed chemical composition (silicate core, sp$^2$ and sp$^3$ carbonaceous layers), and an additional molecular component, in the form of free-flying polycyclic aromatic hydrocarbons. We fit most of the observed extinction profiles. Failures occur for lines of sight presenting remarkable rises blueward the bump. We find a tendency in the carbon chemical structure to become more aliphatic with the galactic activity, and to some extent with increasing redshifts. Moreover, the contribution of the molecular component to the total extinction is more important in younger objects. The results  of the fitting procedure (either successes and failures) may be naturally interpreted through an evolutionary prescription based on the carbon cycle in the interstellar medium of galaxies. 
\end{abstract}
\phantom{A}
\vspace{1cm}

\section{Introduction}
A powerful way of studying dust properties, and in principle to discriminate among different dust production sources, is through the extinction of stellar light as a function of wavelength. Accurate measurements of InterStellar Extinction Curves (ISECs) are almost exclusively limited to our Galaxy (e.g., \citealt{FM07}, and galaxies in the Local Group (e.g., \citealt{C15}) because at greater distances it becomes impossible to obtain the photometry or spectroscopy of individual stars needed for extinction determinations. Derivations of interstellar dust properties of the interstellar medium in high redshift galaxies focus on the study of either the whole galaxy, or within absorption systems along a single line of sight to distant quasars. In the first case dust extinction is inferred from composite emission spectra, and it is the result of blended events of gas and dust absorption, emission and scattering. Thus, the resultant observed stellar dimming is not produced by a true extinction law, but rather by a dust attenuation profile that is highly dependent on the geometric distribution of the dust, gas and stars (e.g., \citealt{C00}). Direct measurements of the interstellar dust properties in high redshift galaxies performed along a single line of sight to quasars typically suffer from the relatively complex spectral shape of these objects, which makes distinguishing dust from intrinsic color variation very challenging (e.g., \citealt{H13}). 

Such difficulties are in part relieved observing the deep and point-like line of sight to a Gamma-Ray Burst (GRB) afterglow. GRBs are intense bursts of extremely high energy observed in distant galaxies, the brightest explosive phenomena in the Universe, followed by a longer-lived afterglow emitted at longer wavelengths from X-rays to the radio domain. Bursts that last more than a couple of seconds are known as long-duration GRBs, and are associated with core-collapse supernova explosions. Long duration GRBs have spectra with simple shapes, basically a featureless broken power-law from the X-ray to the near infrared, with a rising and decaying behavior (e.g., \citealt{GAO15}). They have high intrinsic brightness, and generally occur in dense, star forming environments, making these events ideal in the study of extinction properties of interstellar dust on a cosmological scale. 

To interpret the extinction curves of GRB hosts many studies assume empirical templates, such as the average galactic ISEC  (e.g., \citealt{S11}). An alternative approach that does not require such a strong assumption has been put forward by \citet{Li08} and \citet{LL10}, who exploited a parametrized  functional form for the normalized (to the visible) extinction, the so-called Drude approach.  The description consists of three separate contributions to far-ultraviolet extinction rise, 217.5~nm bump, and near-infrared/visible extinction, respectively, shaped by four dimensionless free parameters. This prescription is similar to the \citet{FM07} parametrization from which differs essentially for the lower number of free parameters, and the more extended (at lower frequencies) domain of applicability. 

One of the major problems in the study of GRB dust is the paucity of photometric or spectroscopic data, that in some cases are outnumbered by model parameters. In those situations \citet{LL10} measure the goodness of their model fit reducing the $\chi^2$ by the number of data (instead of the free parameters). This is conducive to possibly incomplete pictures of the observables. However, such an unsatisfactory situation is alleviated by the \emph{a priori} choice of the extinction profiles, either templates (less number of parameters, but morphologically stiff) and empirical parametrizations (more flexibility). It is unfortunate that the single parameter \citet{CCM} recipe is only valid for the Milky Way Galaxy (MWG), failing when applied even to our closest neighbourhoods \citep{G03}.  

The sparsity of data implies that, given the complexity of the underlying physics and chemistry, any physical dust forward model totally overfits the data, and therefore it cannot be applied directly to the observations. Thus, we assume the extinction profile derived by \citet{LL10} using the empirical description inferred via the Drude approach, and (try to) unfold such synthetic description of dust into physically well-grounded properties. The validity of the physical properties we obtain therefore obviously relies on the validity of the assumed Drude model.

The sample of GRB host galaxies considered in the \citet{LL10} study are located at redshifts $z \la 2$. Then, the use of both templates and empirical representations relies on the reasonable assumption that the nature of dust has not changed that much in the last 10~Gyr. In fact, gathering together data collected in the last decades, it is evident that, whatever the method exploited to derive dust attenuation, the amount of dust in the Universe increases from the earliest epochs to reach a maximum at $z \sim 1 - 1.5$ \citep{B12}. At about $z \sim 3.5$ the dust content reaches about the same level as measured in the local Universe. Beyond $z = 4$ the dust presence fades with increasing redshifts. The dust attenuation peak appears to be delayed by approximately $2-3$ Gyr with respect to the cosmic Star Formation Rate (SFR) density (e.g., \citealt{B10}). A natural way to reconcile the epoch of the maximum SFR with the dust attenuation peak is to suppose that dust is produced by intermediate-mass long-lived stars in the redshift range $z \sim 1.5 - 3$, linked with the fast drop of dust at $z < 1$ because of the decline of new star formation from $z\sim 2$. Even allowing for a net positive contribution from supernovae to the dust budget (see e.g., \citealt{M15}), we still expect interstellar dust to be mostly composed of the familiar heavy elements, i.e. "silicates" and "carbons" \citep{D03}. How these materials are assembled in a grain is still a matter of debate (see \citealt{CP12} for a brief review). Nevertheless, when re-read in physical terms, a threefold representation such as the one put forward by \citet{LL10} ~\textendash~ or \citet{FM07} ~\textendash~ may be interpreted as the superposition of a population of "classical" grains and a (macro-)molecular component of free-flying PAHs (see \citealt{M13}), contributing to both bump ($\pi^\star \leftarrow \pi$ transitions) and far-ultraviolet rise (low energy side of $\sigma^\star \leftarrow \sigma$  transitions). 

Our purpose in this work is to explore the sensitivity of normalized ISEC shapes observed during GRB events to the relative abundances of dust grain components, using a dust model in which the various component contributing to the interstellar extinction are related and respond to the local physical and chemical conditions in the interstellar medium of galaxies. In Section \ref{EDM} we present data and modelling procedure, in Section \ref{Res} we show and comment our results, and finally the conclusions are reported in the last Section. 

\section{Extinction data and modelling}\label{EDM}
The extinction data are a collection of empirical ISECs derived by the analysis of GRB afterglows by \citet{LL10}. These authors exploit a parametrization recipe incorporating the hallmarks of the extinction curves of the MWG, and the Large and Small Magellanic Clouds (LMC, SMC). Such approach is based on a simple formula consisting of four dimensionless parameters embedded in three wavelength-dependent profiles, representing the bump and far-ultraviolet rise floating on the top of a flat, saturated extinction baseline declining redward beyond the visible. The inferred extinction curves show a wide diversity of shapes, ranging from relatively flat curves to extinction profiles resembling those of the MWG, LMC, and SMC. Some are featureless and steeply rise toward the far-ultraviolet (with varying degree of steepness), including flat extinction profiles similar to the so-called "Calzetti" attenuation law of starburst galaxies. Others are galactic-like but with damped 217.5~nm bumps. Extinction "markers" such as the shapes of the inferred ISECs, and the presence or absence of the bump do not show any clear link to the redshifts. 
\begin{figure}
\includegraphics[width=\hsize]{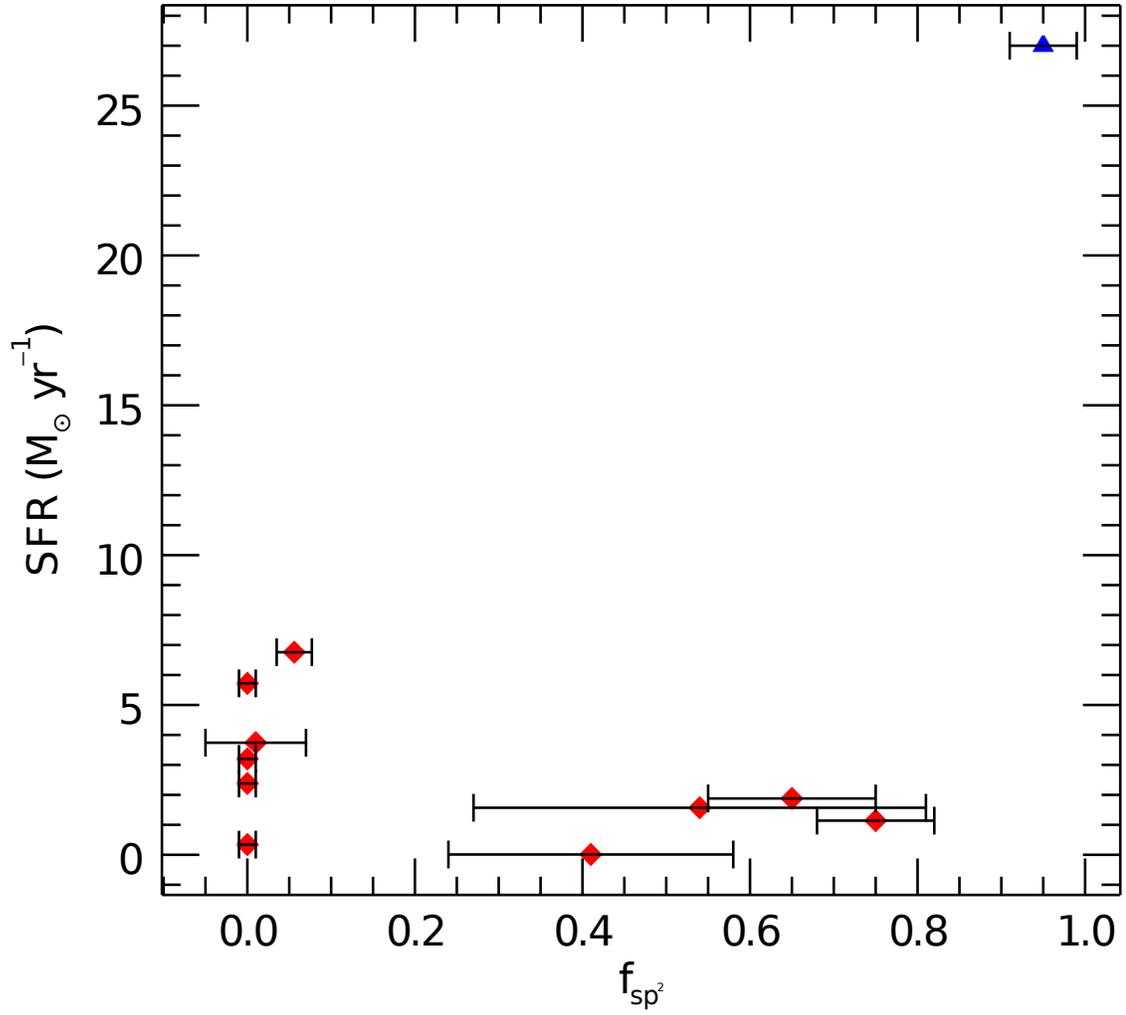}
\caption{The SFR ($M_\odot$~yr$^{-1}$) vs. the mantle aromatic carbon fraction. The blue triangle indicates the (outlier) line of sight GRB061121.}
\label{one}
\end{figure}

The application of a model consisting in a mixture of separate silicate and graphite grains (\citealt{M77}, see \citealt{LL10} for details) do not provide evidence for the evolution of the dust with redshifts, and then with the cosmic epoch. However, a curve-fitting approach does not discriminate well between the various potential dust models considered by \citet{M77}. To this aim we need a dust model specifically constructed on the assumption that dust grains evolve in space. 
\begin{figure}
\includegraphics[width=\hsize]{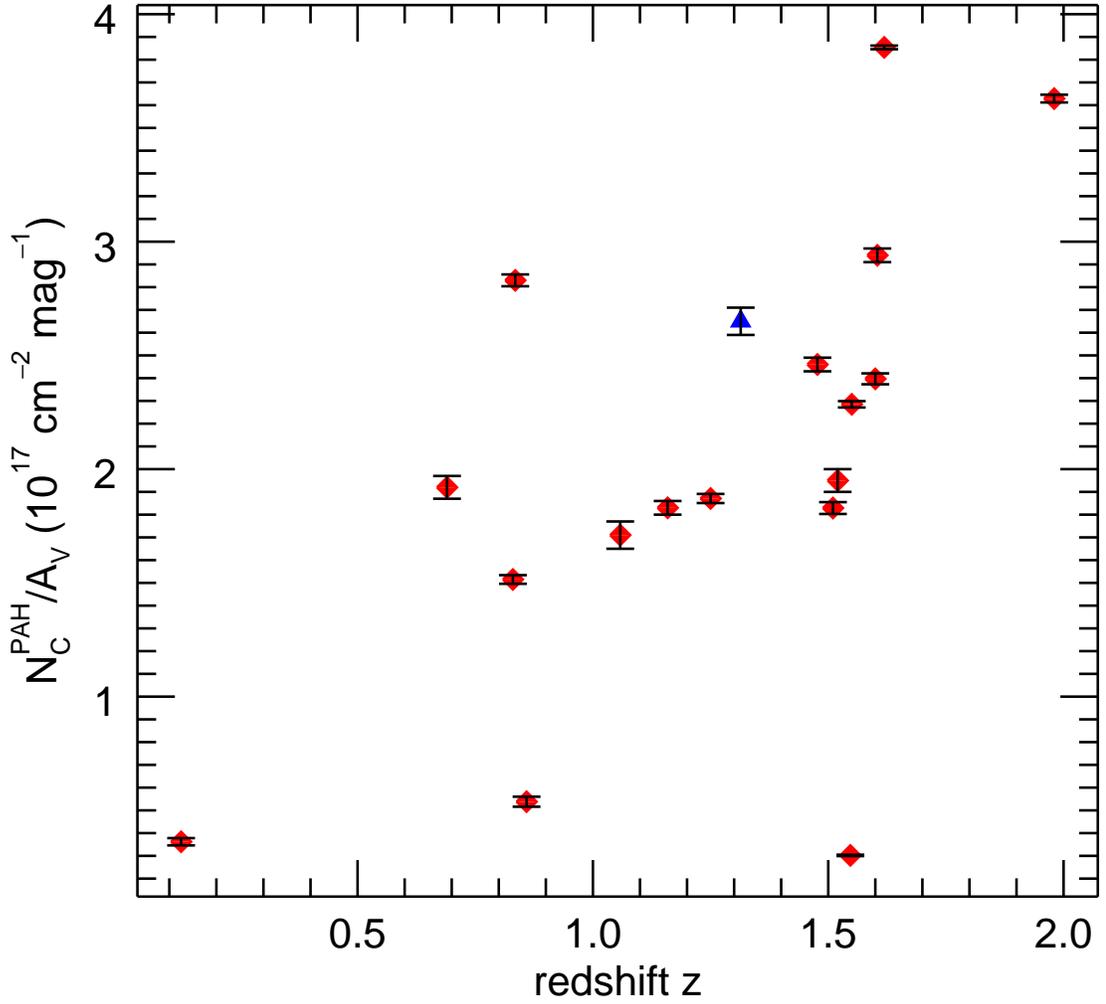}
\caption{Correlation between the carbon column density locked in PAHs normalized to the visual extinction and the redshift. The blue triangle indicates the line of sight GRB061121.}
\label{two}
\end{figure}

Dust grains are formed in the envelopes of cool stars and in novae and supernovae, and are destroyed in shocks. From considerations of the relative responses to shocks of silicate and carbon materials (see e.g.,  \citealt{B14}) we consider model grains consisting of multilayered carbon mantles containing silicate cores \citep{I08}. The model is inherently time-dependent with its optical properties modified by the interaction with the environments in which grains are embedded \citep{CP10}. Characteristic timescales are the deposition time of carbon in a hydrogen-rich gas (mainly sp$^3$ polymeric bondings),  and the time for its conversion to aromatic carbon under the "photo-darkening" action of the local ultraviolet radiation field. In the MWG  typical values are $\sim 10^4 - 10^5$ and $10^5 - 10^6$~yr (e.g., \citealt{CPW98}), respectively. The photo-darkening time may be extended considering the inverse process in which the aromatic component, heated in a hydrogen plasma (e.g., shocks), may rearrange some of its carbon skeletons from sp$^2$ to sp$^3$ bondings. The combined effect is the formation of  an inner sp$^2$ rich layer with new aliphatic material continuously deposited on top; there is also another timescale ~\textendash~ in the MWG generally much longer \citep[about a few million years,][]{CP14} than the other timescales  ~\textendash~ associated with the rapid and partial removal of the carbon layers and recycling in the gas-phase when the dust grains pass through a shock. 

The whole process providing the "instantaneous" grain model needed for the fitting procedure is summarized in a set of differential equations describing carbon deposition onto the silicate cores and its following photo-darkening \citep{CP14}. Fitting techniques and the determination of the observationally based inferences of dust physical parameters are detailed in \citet{M13}. Here, we just recall that individual dust grains are defined by 4 parameters, the void fraction in percentage of the core volume $f_v$ (independent of the core size) simulating core porosity, the core radius $a$, the mantle thickness $w$, and the sp$^2$ fractional content in the mantle $f_{\rm sp^2}$ ($f_{\rm sp^3} = 1-f_{\rm sp^2}$). The corresponding extinction and scattering cross-sections, computed using light scattering techniques (see \citealt{B07}) are connected through the power-law $(a+w)^{-q}$ defining the size distribution of the collection of particles. Such distribution is allowed for a gap, so that two populations of dust grains, big and small ones, may be present, each characterized by lower ($a_-$, $b_-$) and upper  ($a_+$, $b_+$)  size limits in the distribution. 54 types of PAHs constitute the molecular component, and these are in neutral, cation, dication and anion forms, with opacities computed \emph{ab initio} by \citet{Ma07}.

After performing a successful fit on one of the original data sample, we perturb it by a random Gaussian noise, and repeat the fit on the perturbed data. The procedure is iterated to obtain synthetic statistics. In some cases, the fit to the perturbed data results in a better match of the original data. This allows the possibility to discover local minima or saddle points.  In such events, the procedure is restarted using the set of  "perturbed" parameters as initial point. The synthetic statistics are then used to directly estimate their covariance matrix. Statistical errors are reported in all Figures and Tables (numbers between parentheses.

Finally (as discussed e.g., in \citealt{M13}) there is not a unique mixture of the PAHs in our sample that best fits a given extinction curve, but manifold PAH mixtures that all fit it very nearly equally well. While it is impossible to constrain the precise composition of PAHs using only the extinction curve, we can instead constrain rather precisely some of the PAH properties as a group, such as e.g., the total column density $N_{\rm C}^{\rm PAH}$ of carbon atoms in all species, the relative contributions of classical dust and PAHs to the modelled ISECs, and the average  charge per carbon atom, $\langle Q \rangle/{\rm [C]} = \sum_i N_i q_i/ \sum_i N_i {\cal N}_{\rm C}^i$ of the molecular ensemble; $q_i$, $N_i$,  and ${\cal N}_{\rm C}^i$ are charge, column density, and number of carbon atoms for the $i-$th PAH molecule, respectively. 
\begin{figure}
\includegraphics[width=\hsize]{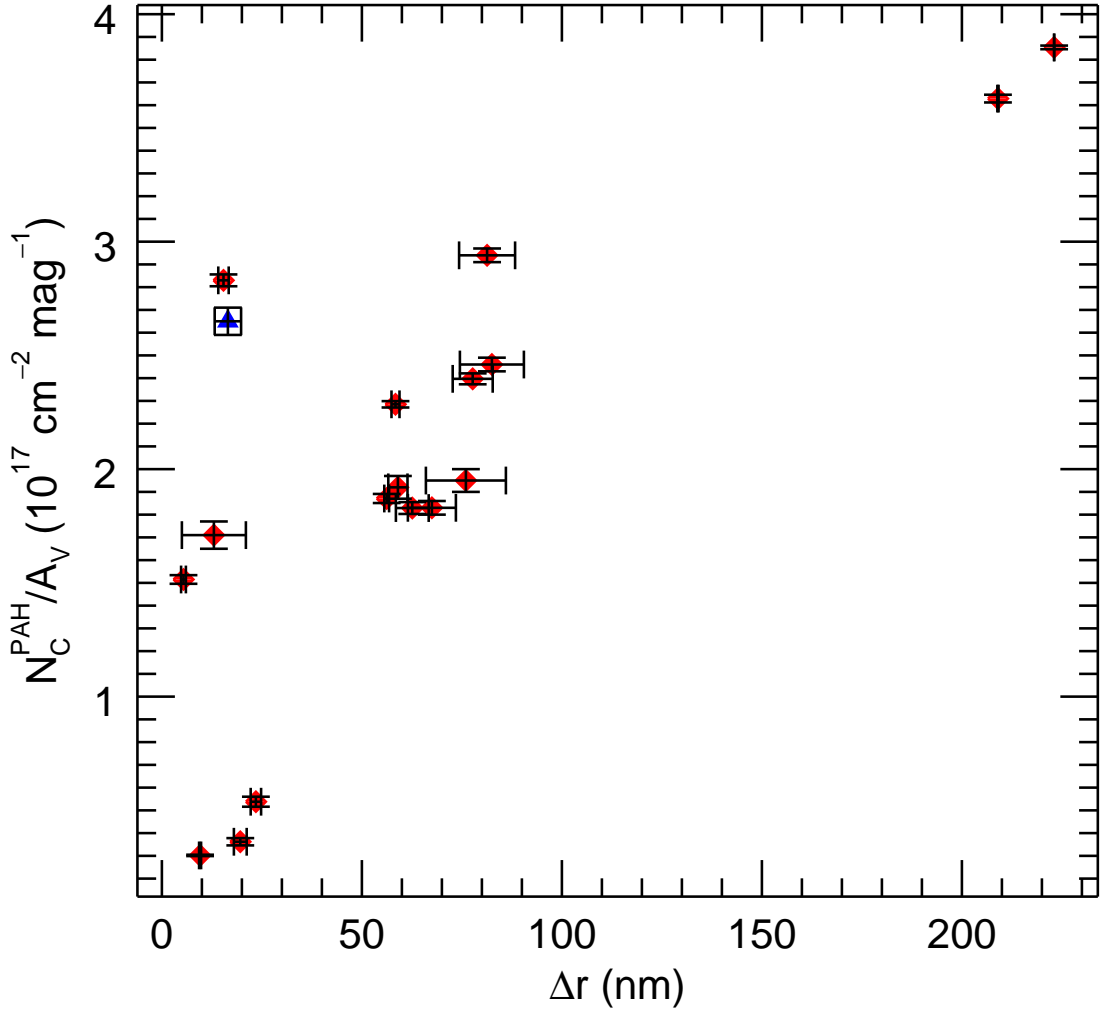}
\caption{Relation between the size gap $\Delta r$ and the carbon column density locked in PAHs normalized to the visual extinction. The blue triangle indicates the line of sight GRB061121.}
\label{three}
\end{figure}

\section{Results}\label{Res}
Applying the dust model outlined in the previous Section (hereafter termed [CM]$^2$) we fit 18 out of 27 lines of sight in the \citet{LL10} sample. In all successful cases the agreement with the data was excellent, meaning that the fitted extinction curve is virtually indistinguishable from the inferred data. Failures occurred for ISECs presenting remarkable rises blueward after the bump. In these cases the normalized extinction in the ultraviolet frequently reached values larger than 10. In 5 cases we could fit the extinction profiles reducing the wavelength domain to $\lambda^{-1} \la 6 - 7$~cm$^{-1}$. Such results are uncertain and we do not report them here. The inferred dust parameters for the fitted ISECs are shown in Tables \ref{tone} (classical dust) and \ref{ttwo} (PAHs). In this latter Table we also state the charge dispersion, $\sigma_Q /[{\rm C}]$, a measure of the homogeneity of the charging within the PAH distribution, while in the last two columns we indicate the redshift and (when available) the SFR in $M_\odot$~yr$^{-1}$. 

\begin{sidewaystable}
\vspace{-1cm}
\caption{Fitting Parameters of the Classical Dust Component for Each Line of Sight$^{(a)}$}
\scriptsize
\hspace{-0.9cm}
\begin{tabular}{ccccccccc}
\hline \hline
LoS & $f_{v}$ & $w$ & $f_{sp^{2}}$ & $a_{-}$ & $a_{+}$ & $b_{-}$ & $b_{+}$ & $q$  \\
\cline{5-8}
        &             & (nm) &                     & \multicolumn{4}{c}{(nm)} & \\
\hline \hline 
GRB000911 & 0.232 (0.020)        & 5.2 (0.6)          & 0.54 (0.27)       & 10.5 (2.9)         & 31 (5) & 44 (6) & 494.7 (0.8) & 3.4890 (0.0022) \\
GRB010222 & 0.189 (0.023)        & 1.47 (0.16)      & 0.0 (0.0) & 5.0 (0.0)          & 44.5 (0.6) & 127 (8) & 493.9 (1.3) & 3.503 (0.007) \\
GRB020405 & 0.148 (0.024)        & 1.97 (0.09)      & 0.01 (0.06)       & 5.0 (0.0) & 49.9 (1.4) & 108.9 (1.9) & 496.9 (1.0) & 3.5341 (0.0021) \\
GRB020813 & 0.220 (0.010)        & 0.530 (0.023)  & 0.056 (0.021)   & 5.0 (0.0) & 56.40 (0.27) & 112.6 (0.5) & 555.6 (1.0) & 3.5604 (0.0014) \\
GRB030226 & 0.2811 (0.0009)     & 0.694 (0.022) & 0.0004 (0.0018) & 5.0 (0.0) & 37.36 (0.05) & 246.37 (0.14) &460.4 (0.8) &3.53250 (0.00018) \\
GRB030328 & 0.273 (0.026)         & 4.3 (0.5)         & 0.0 (0.0)  & 5.0 (0.0) & 53.1 (1.5) &129.1 (10) & 490.2 (1.7) &3.483 (0.006)\\
GRB040924 & 0.252 (0.007)         & 1.12 (0.13)     & 0.65 (0.10)        &5.0 (0.0) & 23.7 (1.1) &47.2 (0.7) &494.13 (0.25) &3.4810 (0.0014)\\
GRB051111 & 0.204 (0.019)          & 0.535 (0.009) & 0.032 (0.010)    &5.0 (0.0) & 70.6 (0.5) &129.0 (0.9) &555.0 (1.3) &3.5564 (0.0021)\\
GRB060614 & 0.292 (0.012)          & 5.43 (0.18)    & 0.41 (0.17)         &5.03 (0.12)      & 30.0 (1.2) &49.6 (1.0) &491.8 (0.6) &3.4587 (0.0012)\\
GRB061121 & 0.21 (0.05)              & 1.58 (0.21)     & 0.95 (0.04)        &5.02 (0.05)      & 26.3 (2.1) &42.8 (2.6) &502.0 (1.0) &3.5054 (0.0024)\\
GRB061126 & 0.263 (0.019)        & 4.2 (0.3)         & 0.0 (0.0)  & 5.0 (0.0) & 53.5 (1.4) &121 (6) &490.9 (1.1)& 3.470 (0.004)\\
GRB070125 & 0.0845 (0.0020)     & 0.730 (0.015)  & 0.0 (0.0)  &10.59 (0.04)    &26.85 (0.08)& 36.40 (0.21)& 553.3 (0.9) &3.5095 (0.0004)\\
GRB071003 & 0.187 (0.022)       &1.26 (0.15)       & 0.0 (0.0)  &5.0 (0.0) & 46.7 (0.8)& 128 (7)& 494.9 (1.3) &3.511 (0.007)\\
GRB080330& 0.243 (0.015)          & 3.86 (0.22)      & 0.0 (0.0)  &5.0 (0.0) & 52.4 (1.1) &115 (4) &492.6 (1.0) &3.483 (0.003)\\
GRB970508 & 0.232 (0.013)         & 5.1 (0.3)          & 0.75 (0.07) &6.69 (0.27)      & 8.3 (0.6) &23.7 (1.2) &499.9 (1.1) &3.4927 (0.0009)\\
GRB990123 & 0.195 (0.018)         &1.42 (0.10)       & 0.0 (0.0)    &5.0 (0.0) & 46.3 (0.7) &124 (5) &493.7 (1.0) &3.496 (0.006)\\
GRB990510 & 0.34103 (0.00025) & 0.701 (0.005)  & 0.0008 (0.0008) &5.0 (0.0) & 34.7065 (0.0021) &257.763 (0.007) &456.16 (0.23) &3.48494 (0.00004)\\
XRF050824X & 0.141 (0.007)       & 0.79 (0.03)       & 0.0 (0.0)  & 5.50 (0.15)       & 27.0 (0.3) &32.4 (0.5) &499.70 (0.29) &3.5031 (0.0012)\\
\hline \hline
\end{tabular}
\flushleft
$(a)$ the statistical errors are shown between parentheses
\label{tone}
\end{sidewaystable}
\begin{sidewaystable}
\centering
\caption{Mean properties of the PAH mixture$^{(a)}$} 
\label{pahtable}
\begin{tabular}{ccccccc}
\\
\hline
\hline
\multicolumn{1}{c}{LoS} & \multicolumn{1}{c}{$N_{\rm C}^{\rm {PAH}}/A_V$}  & \multicolumn{1}{c}{$\langle Q \rangle/[{\rm C}]$}&  \multicolumn{1}{c}{ $\sigma_Q /[{\rm C}]$} & $N_{\rm C}^{\rm {PAH}}$ & $z^{(b)}$ & SFR$^{(c)}$ \\
& \multicolumn{1}{c}{ $\times 10^{17}$ (cm$^{-2}$ mag$^{-1}$)} & \multicolumn{1}{c}{ ($e^-$)} & \multicolumn{1}{c}{ $(e^-)$} & $\times 10^{15}$~(cm$^{-2}$) & & $M_\odot$~yr$^{-1}$\\
\hline \hline
GRB000911 & 1.71 (0.06) & 0.028 (0.006) & 0.063 (0.008) & 54.6 (2.1) & 1.0585 & 1.57 \\
GRB010222 & 2.46 (0.03) & 0.064 (0.003) & 0.053 (0.005) & 71.4 (1.0) & 1.4800 & 0.34 \\
GRB020405 & 1.92 (0.05) & 0.003 (0.005) & 0.0770 (0.0024) & 138 (4) & 0.6910 & 3.74 \\
GRB020813 & 1.871 (0.020) & -0.0232 (0.0027) & 0.0587 (0.0021) & 63.6 (0.7) & 1.2550 &6.76 \\
GRB030226 & 3.629 (0.017) & 0.0370 (0.0016) & 0.0656 (0.0016) & 87.1 (0.4) & 1.9860	&\\
GRB030328 & 1.95 (0.05) & 0.0302 (0.0029) & 0.083 (0.003) & 39.0 (1.0) & 1.5200 & 3.20\\
GRB040924 & 0.538 (0.022) & 0.072 (0.005) & 0.049 (0.009) & 19.4 (0.8) & 0.8590 & 1.88 \\
GRB051111 & 2.285 (0.014) & -0.0459 (0.0011) & 0.0430 (0.0021) & 86.8 (0.5)& 1.5500 & \\
GRB060614 & 0.362 (0.016) & 0.005 (0.008) & 0.076 (0.004) & 15.2 (0.7)& 0.1250 & 0.01\\
GRB061121 & 2.65 (0.06) & 0.039 (0.004) & 0.049 (0.005) & 292 (7)& 1.3145 & 27.0\\
GRB061126 & 1.83 (0.03) & 0.0299 (0.0028) & 0.083 (0.003) & 5.48 (0.10)& 1.1588 & 2.38 \\
GRB070125 & 0.302 (0.003) & -0.051 (0.004) & 0.039 (0.007) & 13.61 (0.14)&1.5471 &\\
GRB071003 & 2.94 (0.03) & 0.0638 (0.0026) & 0.049 (0.004) & 150.0 (1.7)&1.60435&\\
GRB080330 & 1.829 (0.026) & 0.028 (0.003) & 0.083 (0.003) & 75.0 (1.0)&1.5119&\\
GRB970508 & 2.830 (0.026) & 0.023 (0.003) & 0.054 (0.003) & 42.5 (0.4)& 0.8350&1.14\\
GRB990123 & 2.397 (0.024) & 0.0565 (0.0027) & 0.063 (0.004) & 71.9 (0.7)&1.6000&5.72\\
GRB990510 & 3.854 (0.008) & 0.0759 (0.0006) & 0.0333 (0.0008) & 142.6 (0.3)&1.6190&\\
XRF050824X & 1.515 (0.019) & -0.0452 (0.0025) & 0.0345 (0.0021) & 34.8 (0.4)&0.8278	&\\
\hline \hline
\end{tabular}
\flushleft
$(a)$ the statistical errors are shown between parentheses \\
$(b)$ taken from www.grbhosts.org with the exception of GRB071003 given in \citet{LL10} \\
$(c)$ taken from www.grbhosts.org
\label{ttwo}
\end{sidewaystable}

The charge dispersion of the molecular components decreases with increasing charge (in absolute value), suggesting that when charge dispersions are low we are sampling localized regions along a line of sight. The carbon column density locked in PAHs normalized to the visual extinction $N_{\rm C}^{\rm {PAH}}/A_V$ correlates with the redshift (Figure~\ref{two}), but not with the SFR. $N_{\rm C}^{\rm {PAH}}/A_V$ also shows a weaker dependence from the classical grain size gap $\Delta r$ (Figure \ref{three}). Finally, the absolute PAH column density is not related to any other quantities.

\section{Discussion and Conclusions}
We analyzed a relatively small sample of lines of sight towards GRBs exploiting the [CM]$^2$ dust model, and we interpret the results of the fitting procedure through an evolutionary prescription based on the carbon cycle in the interstellar medium of galaxies. The novel aspect of this work is that in the dust model used, the PAHs and other atoms and fragments of erosion are part of the natural circulation of carbon in the interstellar medium between gas and solid phases. In this latter phase the mantle thickness and its chemical composition are determined by the local physical conditions. It appears that all known types of observed ISECs can be accounted for on the basis of the [CM]$^2$ model (see e.g., \citealt{Z11}).
\begin{figure}
\includegraphics[width=\hsize]{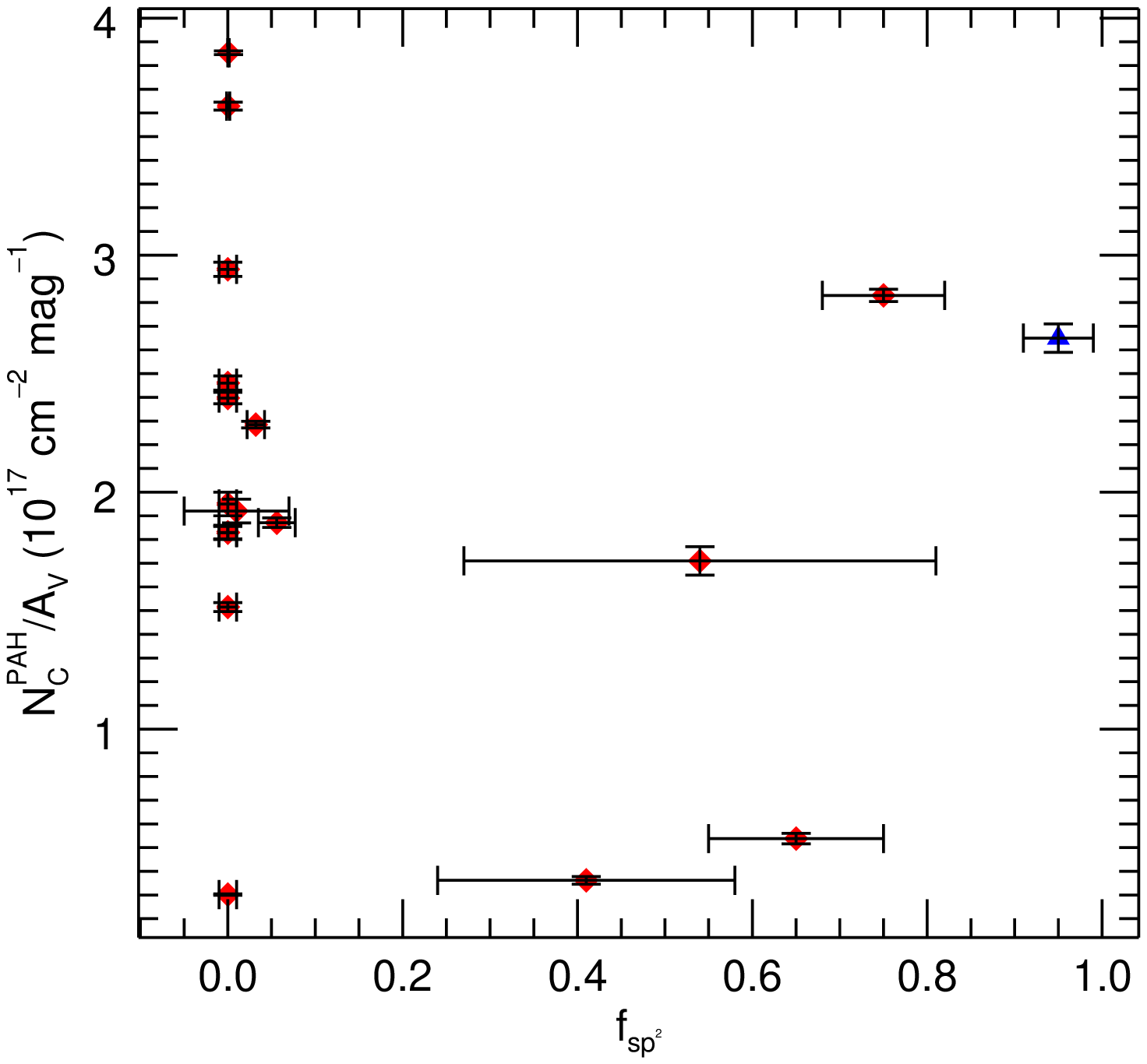}
\caption{Relation between the mantle aromatic fraction $f_{sp^{2}}$ and the carbon column density locked in PAHs normalized to the visual extinction. The blue triangle indicates the line of sight GRB061121.}
\label{four}
\end{figure}

The results of the fitting procedure shown in the preceding Section may be naturally interpreted in the framework of the carbon cycle outlined above. The results shown in Figure \ref{one} seem to suggest that carbonaceous mantles are increasingly aliphatic with increasing SFR considered as a reliable measure of the galactic activity. Thus, the mantle annealing timescale must be almost always much longer than the average time between mantle shattering events. The annealing time is defined by the competition between energetic processing by ultraviolet radiation and re-hydrogenation by hot H atomic gas, turning aromatic carbonaceous material back into aliphatic. In the MWG, re-hydrogenation was deemed to be by far negligible, but this appears not to be the case in the majority of lines of sight towards the GRB events considered here. Since the SFR increases with increasing redshifts (in the range of redshifts considered in this study), "far" dust loses the aromatic character pervading quiescent galaxies such as the MWG. 

It is important to realize, however, that dust properties determined using GRBs are not totally representative of the conditions in the host galaxies. Moreover, sometimes host associations are uncertain (e.g., \citealt{H10}). Although the considered lines of sight appears to be sufficiently typical to provide correlation with global galactic properties, this is not possibly true for the line of sight towards the event GRB061121, in which the most MWG-like ISEC within the sample occurs. The prevalent aromatization of dust mantles and the abundance in free-flying PAHs (with respect to the visual extinction) are not consistent with the high SFR~$=  27~M_\odot$~yr$^{-1}$ present in the host galaxy. This discrepancy may be alleviated decreasing the sp$^3 \to$~sp$^2$ conversion time, in competition with re-hydrogenation induced by the large activity of the galaxy as summarized by its SFR. The shape of the ISEC is unaffected as long as the ratio between collisional and photo-darkening rates has kept unchanged.  Collisional rates are proportional to the product $n_{\rm H} T_k^{1/2}$, while the annealing rate to the intensity of the local radiation field 
$\cal{X}$ (e.g., in units of the Habing's field). Thus, the increase of $\cal{X}$ by a factor e.g., hundred is balanced by a corresponding increase in the gas pressure (incidentally consistent with the required  larger radiation energy deposition). The ISEC towards GRB061121 appears to be then consistent with the presence of a photon-dominated region, not at all incongruous since GRB events are often associated with star-forming regions. It is worthwhile to note that GRB061121 presents the lowest charge dispersion and one of the highest positive charge of the PAH mixture.   

More interesting is the well-defined increase of $N_{\rm C}^{\rm {PAH}}/A_V$ with the redshift $z$. Such increase should be read in relative terms, as a major contribution of PAHs with respect to the classical dust grains. If we compare the aromatization fraction of mantles with the fractional concentration of PAHs (Figure \ref{four}) we note the existence of two separate regimes, one characterized by $f_{sp^{2}}$ close to zero, the other one associated with positive $f_{sp^{2}}$ values. In the first case, it is not apparent any clear correlation, while in the second case $f_{sp^{2}}$ grows with $N_{\rm C}^{\rm {PAH}}/A_V$. This latter trends is characteristic of low activity galaxies, as supported by the anticorrelation of $f_{sp^{2}}$ and SFRs. In the first case, the situation is reminiscent of the environments of the LMC and SMC, i.e. the carbon recycling time must be much shorter than the local 
photo-darkening time. 

From a physical point of view the uncorrelation of $f_{sp^{2}}$ with the PAH concentration evidences  a potential problem in our description of the carbon cycle in galaxies posed by the sharp decline of the source of aromatic material in the diffuse interstellar medium, since most dust mantles are destroyed before they experience substantial annealing. Since PAH formation rates in the cold, carbon-rich winds of evolved stars are too slow, the injection time being approximately 2 Gyr, such component must grow in the diffuse medium or be formed by another yet unknown mechanism. Such a mechanism may be connected to the fate of the plethora of aliphatic fragments released by shattering events. They could evaporate into polyyines, and thereafter quickly photodissociated as proposed by \citet{DW84}, or survive as a population of nanoparticles for a significant time contributing trough $\sigma^\star \leftarrow \sigma$  transitions to the far-ultraviolet rise (but not to the bump). This additional component, missing in the current version of the [CM]$^2$ model may recover the failures in those lines of sight exhibiting extraordinary far-ultraviolet rises. Over time, depending on the competition of collisional (destruction) and radiative (structural transformation) events this population might be (partially) converted into PAHs.

The lack of correlation of the 217.5~nm bump with redshifts evidenced by \citet{LL10} may be easily explained by the  combined action of the classic and molecular components. In other words within the [CM]$^2$ model an increase in the PAH column density does not translate straightforwardly in an increase of the bump intensity.

In conclusion we model the ISECs inferred along a sample of lines of sight to GRB afterglows with a synthetic population of dust grains consisting of core-mantle particles and a collection of free-flying PAHs, providing excellent fits. While this result is not particularly interesting, the retrieval of dust physical properties through the application of the [CM]$^2$ forward model produces a solid base on which to discuss the nature of dust in the local and distant Universe. The major results of these work are: (1) there is tendency in the chemical structure of carbon dust to become more aliphatic with the galactic activity, and to some extent with increasing redshifts; (2) the contribution of the molecular component (PAHs) to the total extinction is more important at early times. Along some lines of sight the lack of any relation between mantle aromatic fractions and relative abundances of PAHs suggest the existence of a  transient aliphatic carbon component.

\begin{acknowledgments}
We acknowledge the support of the Autonomous Region of Sardinia, Project CRP 26666 (Regional Law 7/2007, Call 2010). This research has made use of the GHostS database (www.grbhosts.org), which is partly funded by Spitzer/NASA grant RSA Agreement No. 1287913.
\end{acknowledgments}

\end{document}